\documentclass[10pt,conference]{IEEEtran}
\usepackage{cite}
\usepackage{amsmath,amssymb,amsfonts}
\usepackage{algorithmic}
\usepackage{graphicx}
\usepackage{textcomp}
\usepackage{xcolor}
\usepackage[hyphens]{url}
\usepackage{fancyhdr}
\usepackage{hyperref}

\makeatletter
\let\MYcaption\@makecaption
\makeatother

\usepackage{subcaption}

\makeatletter
\let\@makecaption\MYcaption
\makeatother

\usepackage{tikz}
\usepackage{multirow}

\definecolor{darkgreen}{rgb}{0.0, 0.5, 0.0}

\newcommand*\circled[1]{\tikz[baseline=(char.base)]{
            \node[shape=circle,draw,inner sep=0.5pt] (char) {#1};}}

\title{Mitigating Classical Resource Costs in Quantum Error Correction via Generalized qLDPC Predecoding}

\newcommand\paperauthors{Alexander Knapen$^\dagger$, Junyi Luo$^\dagger$, Guanchen Tao$^\dagger$, Yuxuan Wang$^\S$, Tomas Bruno$^\dagger$,\\Qirui Zhang$^\dagger$, Dennis Sylvester$^\dagger$, Mehdi Saligane$^\ddagger$, and Gokul Subramanian Ravi$^\dagger$}
\newcommand\affiliations{$^\dagger$University of Michigan, Ann Arbor; $^\ddagger$Brown University; $^\S$University of Pennsylvania}
\newcommand\emails{\\Email(s): \{aknapen, guanchen, tbruno, dmcs, qiruizh, gsravi\}@umich.edu,\\ \{junyi\_luo, mehdi\_saligane\}@brown.edu,\\wayuxua@engineering.upenn.edu}

\author{
    \IEEEauthorblockN{\paperauthors{}}
    \vspace{0.2cm}
      \IEEEauthorblockA{
        \affiliations{}
        \vspace{0.2cm}
        \emails{}
      }
}

\begin{document}
\maketitle

\thispagestyle{plain}
\pagestyle{plain}

\begin{abstract}
    Large-scale fault-tolerant quantum computing (FTQC) will require quantum-classical interfaces (QCIs) that orchestrate real-time decoding over thousands to millions of logical qubits simultaneously. To scale FTQC systems, complex decoding resources must be shared between logical qubits, creating resource contention bottlenecks in the QCI. Mitigating this contention via optimal resource allocation remains an open problem. Lightweight predecoding techniques can reduce decoder utilization and average latency, both of which ease contention for shared decoding resources. To date, both decoder allocation and predecoding work is limited to the surface code. As focus shifts towards general qLDPC codes, slower decoding exacerbates resource contention, while code complexity precludes manual predecoder design. 

    To address this gap, we introduce an automated framework designed to generate predecoders for arbitrary qLDPC codes. By independently handling up to 99.98\% of the decoding workload, these predecoders reduce decoder utilization up to 4,090$\times$, including up to 81.19\% decrease in expensive OSD post-processing and 59.96\% decrease in extra RelayBP legs. An efficient, pipelined hardware architecture enables simultaneous decoding of $\sim$1,800 BB code logical qubits on a single FPGA, while cryogenic ASIC implementation supports $\sim$50,000-500,000 BB code logical qubits within a 1.5 W power budget at 4 K.
\end{abstract}
\section{Introduction} \label{sec:introduction}
\begin{figure}
    \centering
    \includegraphics[width=\columnwidth]{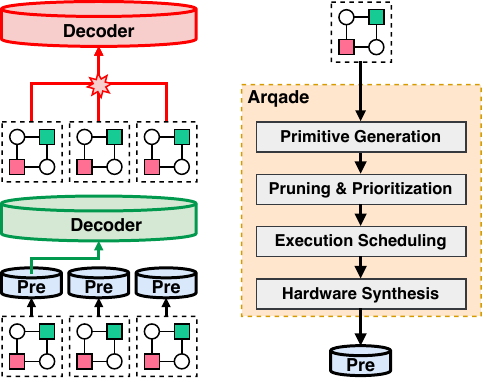}
    \caption{(Left) Multiple logical qubits (dashed squares), implemented in any qLDPC code, contend for the same decoding resources. (Right) Our solution, Arqade, automatically constructs lightweight hardware predecoders to drastically reduce the utilization of decoders and minimize resource contention.}
    \label{fig:overview}
\end{figure}

Quantum error correction (QEC) is a fundamental requirement for transitioning from today's noisy, intermediate-scale quantum (NISQ) computers ($\sim$$10^3$ qubits, $\sim$$10^{-3}$ error rates \cite{acharya2024quantum}) to fault-tolerant quantum computing (FTQC) systems ($\sim$$10^6$ qubits, $\sim$$10^{-15}$ error rates \cite{beverland2022assessing}) capable of solving useful problems. Like classical error correction, QEC uses redundancy to encode some number of robust \textit{logical qubits} using a larger number of noisy \textit{physical qubits} in hardware. How these logical qubits are encoded defines the QEC code.

While many QEC codes have been proposed, quantum low-density parity check (qLDPC) codes \cite{breuckmann2021quantum} are the most promising and well-studied candidates to date. In qLDPC codes, each qubit interacts with only a small, constant number of other qubits which simplifies their implementation in hardware. Within this broad family, the surface code \cite{dennis2002topological, fowler2012surface}, color code \cite{Bombin_2006}, and bivariate bicycle (BB) codes \cite{bravyi2024high} have emerged as attractive candidates, each with unique benefits and drawbacks.

Regardless of the chosen qLDPC code, several challenges must be overcome at the quantum-classical interface (QCI) to realize their implementation in practice. Within the classical control system, a decoder is used to correct qubit errors. Not only must this decoder be extremely accurate to achieve low logical error rates, but it must also process data in real-time with very low latency (e.g., $\sim$1 $\mu s$ for superconducting qubits \cite{skoric2023parallel}). Unlike the surface code, which benefits from fast decoders like minimum-weight-perfect matching (MWPM) \cite{edmonds1965paths, higgott2025sparse, wu2023fusion, wu2025micro} and Union-Find \cite{delfosse2021almost, liyanage2023scalable}, general qLDPC codes rely on belief propagation (BP) \cite{poulin2008iterative, panteleev2021degenerate, hillmann2025localized, yao2024belief, muller2025improved} which suffers from slow convergence and requires expensive post-processing techniques like ordered statistics decoding (OSD) \cite{panteleev2021degenerate} or sequential chains of BP instances \cite{muller2025improved} to maintain high accuracy.

While parallel window decoding \cite{skoric2023parallel, tan2023scalable} relaxes latency constraints, it requires multiple decoders per logical qubit \cite{maurya2024managing} which is not scalable in FTQC systems requiring simultaneous decoding of thousands to millions of logical qubits. Hence, decoders will need to be shared among multiple logical qubits, creating contention for access to these shared resources at the QCI (left, Fig. \ref{fig:overview}). Managing this contention via optimal decoder resource allocation constitutes a classical systems problem, rather than a pure quantum one.

The data used by decoders to infer corrections is generated via syndrome measurement (SM) circuits which measure the code's check qubits. In real-time FTQC systems, these circuits generate large amounts of data at very high rates, necessitating a QCI composed of high-bandwidth interconnects between the quantum chip and the classical control system. This is particularly daunting to engineer for systems with cryogenic qubits, where transmission power budgets are limited and the quantum-classical interface hardware are separated by large spatial and thermal gaps. 

To address these challenges, hierarchical predecoding has been proposed \cite{delfosse2020hierarchical, ravi2023better, chamberland2023techniques, smith2023local, alavisamani2024promatch, knapen2026pinball} which uses a lightweight, locally accurate first-layer predecoder to simplify error data before propagating it to a traditional decoder in the second layer. Depending on their implementation, predecoders can reduce either average decoding latency \cite{alavisamani2024promatch, chamberland2023techniques, smith2023local} or QEC bandwidth requirements \cite{delfosse2020hierarchical, ravi2023better, knapen2026pinball} in the QCI. However, predecoders have only ever been demonstrated for the surface code, leaving qLDPC codes vulnerable to bandwidth, latency, and resource contention bottlenecks.

This naturally motivates the development of predecoders for more general qLDPC codes. However, the sheer number of qLDPC codes, all with distinct properties and structure, renders traditional, manual construction methods intractable. Additionally, since many qLDPC codes require higher-weight and long-range qubit interactions, error propagation under realistic noise becomes more complex and geometrically non-local. Consequently, it not clear whether sufficient error sparsity, the key property underpinning the design of all predecoders, even holds in general for qLDPC codes.

To address these complexities, this paper introduces Arqade\footnote{We use the name ``Arqade" to refer to both the framework used to generate the predecoders as well as to the predecoders themselves.}, a framework to automatically construct predecoders for general qLDPC codes (right, Fig. \ref{fig:overview}). Given any valid implementation of a qLDPC code's SM circuit, Arqade abstracts predecoders into sets of lightweight units of logic called predecoding primitives which can cover $>$95\% of all errors. Using code-agnostic design principles executed over the code's detector error model \cite{gidney2021stim}, Arqade generates all necessary primitives and prunes them into a final, irreducible set. 

To prevent data hazards among predecoding primitives (see Sec. \ref{sec:hardware-arch}), Arqade automatically groups primitives into maximal, non-conflicting groups ordered by an execution prioritization scheme. It synthesizes this output into resource-efficient FPGA or 4 K-compatible ASIC hardware pipelines. Pipeline stage removal targeting less commonly used primitives further reduces power and area without sacrificing accuracy.

When combined with BP-OSD or RelayBP, two of the most accurate and widely tested decoders across diverse qLDPC codes, Arqade predecoders significantly reduce second-level decoder utilization, average decoding latency, and QEC data bandwidth requirements, all of which reduce QCI design complexity, simplify system-wide decoder allocation, and enable scalability into the FTQC regime. In summary, this paper's main contributions are:

\begin{itemize}
    \item \textbf{Automated qLDPC Predecoder Construction:} We develop, Arqade, a framework which uses a qLDPC code's SM circuit to build a predecoder tailored to that code. Using the SM circuit's detector error model, Arqade automatically constructs predecoding primitives for length-1 errors, applying pruning strategies to refine them it into an irreducible set that minimizes design complexity.
    \item \textbf{System-Level Benefits Across Codes:} We demonstrate Arqade's extensibility, evaluating it over eleven codes from five different code families. Across these codes, Arqade achieves up to 99.98\% coverage of the decoding workload, decreasing decoder utilization up to 4,090$\times$. When paired with BP-OSD, it mitigates up to 81.19\% of OSD utilization, and when paired with RelayBP, it achieves 59.96\% reduction in extra sequential relay legs, in both cases reducing contention and shortening average decoding latency.
    \item \textbf{Efficient Hardware Architecture:} To avoid race conditions between predecoding primitives attempting to apply corrections using shared syndrome values, we formulate a novel, qLDPC-inspired graph coloring algorithm that Arqade solves to organize them into non-conflicting groups while maximizing concurrency. It translates these groups into a minimum-depth hardware pipeline amenable to both FPGAs and ASICs. We find that Arqade supports up to 1,800 BB code logical qubits on a single ZCU102 FPGA \cite{amd2023zcu}, while cryogenic ASIC implementation supports $\sim$50,000-500,000 BB code logical qubits within the power budgets at the 4 K stage of a dilution refrigerator.
    \item \textbf{Resource-Constrained Design Optimization:} To enable Arqade's utility under severe resource constraints, we demonstrate that pipeline stages can be progressively removed to further reduce area and power overheads. Importantly, this optimization affects only coverage, not logical error rate, showcasing the opportunity for fine-grained tradeoff analysis and design space exploration. Up to 32\% (39.24\%) additional area (power) savings are possible at a maximum coverage loss of just 4.66\%.
\end{itemize}

\section{Background and Motivation} \label{sec:background}
\subsection{Quantum Error Correction}
The goal of quantum error correction (QEC) is to encode multiple, faulty physical qubits in hardware into logical qubits with much lower error rates. For many QEC codes, if physical error rates can be kept below a certain threshold, then logical errors can be exponentially suppressed.

Codes are represented using the $[[n,k,d]]$ notation, where $n$ physical qubits are used to encode $k$ logical qubits with code distance $d$. The \textit{code distance} is the minimum number of errors needed to move from one logical state to another. Increasing a code's distance increases logical qubit reliability at the cost of more physical qubits. The \textit{encoding rate}, $\frac{k}{n}$ measures how efficiently a code can be implemented.

\subsubsection{qLDPC Codes}
Among the many QEC code families \cite{ErrorCorrectionZoo}, quantum low-density parity check (qLDPC) codes \cite{breuckmann2021quantum} are some of the most promising and well-studied to date. qLDPC codes use two types of physical qubits: \textit{data qubits}, which encode the logical state, and \textit{check qubits}, which detect errors on the data qubits. Their key advantage is that each data (check) qubit interacts with only a small, constant number of check (data) qubits. 

Stabilizer-based qLDPC codes \cite{gottesman1997stabilizer} are particularly noteworthy. They are defined by sets of stabilizers, or Pauli operators that act trivially on the encoded state. Errors on data qubits anti-commute with one or more of these stabilizers, allowing check qubits to measure the stabilizers to detect errors. The discovery of such codes, along with associated logical operators \cite{cohen2022low}, has led to many proposals and small-scale demonstrations of their implementation in hardware \cite{bravyi2024high, lacroix2025scaling, mathews2025placing, rosenfeld2025magic, acharya2024quantum, wang2026demonstration}. Prominent examples include surface \cite{dennis2002topological, fowler2012surface}, color \cite{Bombin_2006}, and bivariate bicycle (BB) codes \cite{bravyi2024high}.

With the continual, rapid evolution of codes and FTQC resource requirements \cite{gidney2021factor, gidney2025factor, webster2026pinnacle, cain2026shor}, it is not known which code or set of codes will be ``best". Thus, while research targeting specific codes and their optimization remains hugely important, research optimizing over wider categories of codes is also critical to mitigate short-term obsolescence. This paper's contributions fall into the latter category.

\subsubsection{Syndrome Measurement Circuits} \label{sec:sm-circuit}
\begin{figure}
    \centering
    \includegraphics[width=\columnwidth]{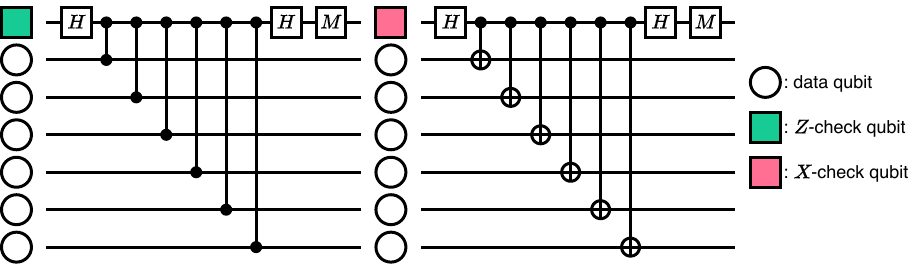}
    \caption{Circuits measuring (left) $XXXXXX$ and (right) $ZZZZZZ$ stabilizers in the BB code.}
    \label{fig:bb_ckt}
\end{figure}

QEC codes are implemented on a quantum device via syndrome measurement (SM) circuits. SM circuits repeatedly entangle check and data qubits such that errors are detected by measuring check qubits. Repeated check measurements form a binary vector called the \textit{syndrome}, where 1s (active) indicate the presence of errors and 0s (inactive) indicate their absence. In a stabilizer code, the SM circuit measures the code's stabilizers; examples for the BB code's weight-6 stabilizers are shown in Fig. \ref{fig:bb_ckt}.

SM circuits are composed of faulty gate and measurement operations that introduce additional errors. Two-qubit gates are particularly problematic, as they introduce hook errors \cite{dennis2002topological} that propagate from a check qubit to multiple data qubits. Furthermore, multiple SM rounds (typically $d$) are needed to combat noisy syndrome measurement values. Depending on the two-qubit gate schedule, an error can flip syndrome values in multiple SM rounds.

\subsection{Decoding} \label{sec:decoding}

The binary syndrome data generated by SM circuits is processed in real-time using a complex decoding algorithm to ascertain whether errors have occurred. In stabilizer codes, it is unnecessary to explicitly identify and correct individual physical errors. Instead, decoders track how accumulated errors affect the outcome of measuring the code's \textit{logical observables}. Decoder corrections then simplify to Pauli frame updates \cite{knill2005quantum} in software that specify how subsequent measurements of these observables should be interpreted. 

Decoders operate on a 3D decoding graph in which nodes represent syndrome values (0 or 1) and edges represent circuit faults that flip the syndrome values at their endpoints. Each two-dimensional slice of the graph represents one round of the SM circuit. While the surface code's simple graph edges allow accurate and efficient \textit{minimum weight perfect matching (MWPM)} decoding \cite{higgott2025sparse, wu2023fusion, wu2025micro}, general qLDPC decoding graphs features hyperedges that instead require iterative \textit{belief propagation (BP)} algorithms \cite{poulin2008iterative, panteleev2021degenerate, hillmann2025localized, yao2024belief, muller2025improved}. Furthermore, maintaining low error rates for these qLDPC codes often necessitates expensive, high-latency post-processing like ordered statistics decoding in BP-OSD \cite{panteleev2021degenerate} or sequential chains of BP instances in RelayBP \cite{muller2025improved}.

\subsection{Quantum-Classical Interface}
The quantum-classical interface (QCI) for FTQC systems will require many high-accuracy decoders processing thousands of logical qubits in real-time to avoid exponential algorithm slowdown \cite{terhal2015quantum} and increases in quantum resources \cite{khalid2025impacts}. For systems with cryogenic qubits, the QCI faces additional power and bandwidth constraints to rapidly transmit syndrome data across temperature stages \cite{das2022afs, knapen2026pinball}. Parallel window decoding \cite{skoric2023parallel, tan2023scalable} can partially relax per-decoder latency requirements, but at the cost of more decoder instances \cite{maurya2024managing}.

Generally, slower decoding algorithms require significant investment in compute resources: efficient MWPM decoding can run on a single CPU core \cite{higgott2025sparse, wu2025micro}, but BP-OSD requires a dedicated GPU cluster \cite{nvidiaRealTime2025} and RelayBP consumes significant, high-end FPGA resources \cite{maurer2025real}. Clearly, to maintain system scalability into FTQC, such decoding resources will need to be shared (e.g., logical qubits sharing BP instances, or BP instances sharing more expensive OSD instances).

Together, resource sharing and slow decoding create contention for decoders in the QCI (Fig. \ref{fig:overview}, left). Thus, intelligent assignment of decoders to logical qubits becomes a classical systems problem of \textit{optimal resource allocation}. Prior work has explored allocating decoders of varying complexity \cite{toshio2025decoder, delfosse2023choose} or dynamically allocating decoders based on varying logical qubit demands \cite{maurya2024managing}, but only in the context of the surface code.

\subsection{Predecoding}
Decoders are complex because, to achieve target logical error rates, they must provision for worst-case error scenarios. In the surface code, the vast majority of error patterns are very sparse (length-1, i.e., spanning one edge of the decoding graph) \cite{ravi2023better} and do not require complex decoding logic to accurately correct. Leveraging this observation, several works have proposed \textit{predecoders} \cite{delfosse2020hierarchical, smith2023local, chamberland2023techniques, ravi2023better, alavisamani2024promatch, knapen2026pinball} which use lightweight logic to pre-process common, length-1 errors. Within this framework, decoding proceeds via a two-level hierarchy, with the predecoder occupying the first level and the `full' decoder occupying the second.

In Ref. \cite{alavisamani2024promatch}, predecoders were divided into two categories: \textit{syndrome-modifying (SM) predecoders} \cite{alavisamani2024promatch, smith2023local, chamberland2023techniques} and \textit{non-syndrome-modifying (NSM) predecoders} \cite{delfosse2020hierarchical, ravi2023better, knapen2026pinball}. SM predecoders sparsify the syndrome data transmitted to the decoder to reduce its average latency. NSM predecoders generate full corrections for common, simple error patterns while detecting rare, complex patterns. Only the latter are propagated to the decoder, reducing the frequency with which it is invoked. Critically, since NSM predecoding reduces second-level decoder utilization, it promises a low-overhead mechanism to mitigate decoding resource contention in the QCI and simplify the decoder allocation problem in FTQC. For cryogenic qubit systems, placing NSM decoders at the 4 K stage of the dilution refrigerator additionally lowers QCI bandwidth and energy consumption \cite{knapen2026pinball}.

\begin{figure}
    \centering
    \includegraphics[width=\columnwidth]{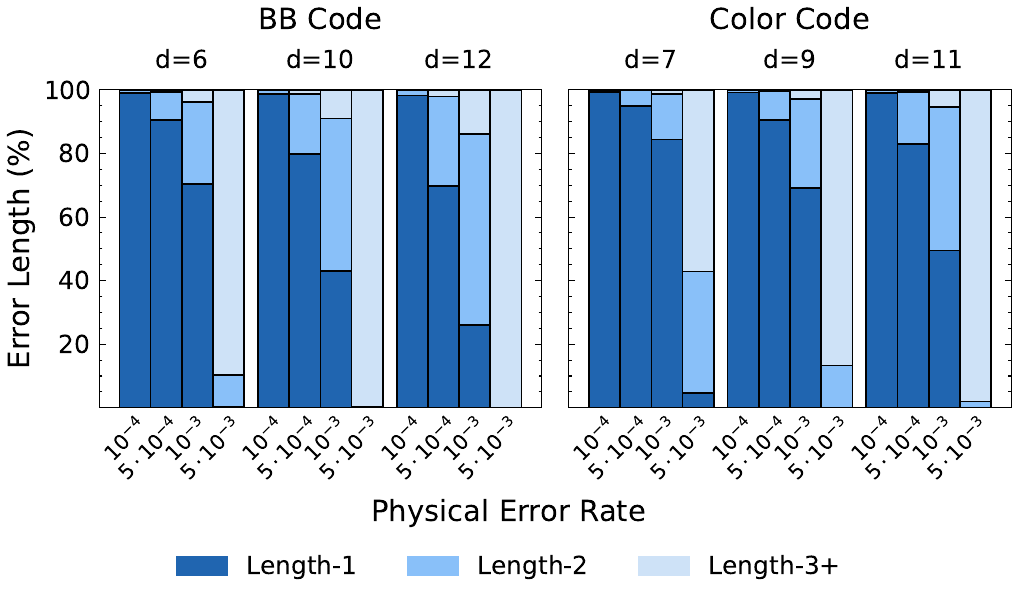}
    \caption{Distribution of maximum-length errors in the BB and color codes.}
    \label{fig:error-length-dist}
    \vspace{-0.2cm}
\end{figure}

\begin{figure}
    \centering
    \begin{subfigure}{0.45\columnwidth}
        \includegraphics[width=\textwidth]{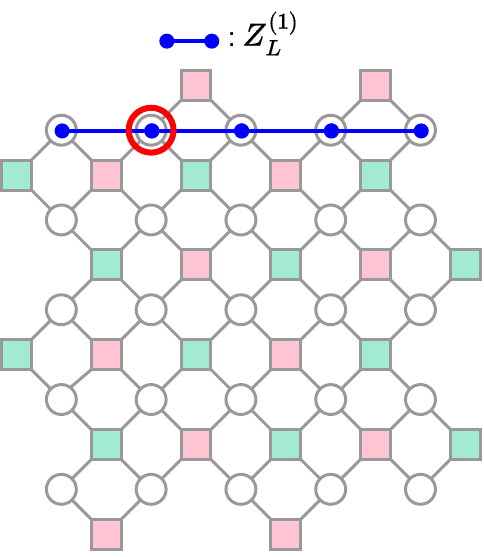}
        \label{fig:logical_sc}
    \end{subfigure}
    \begin{subfigure}{0.45\columnwidth}
        \includegraphics[width=\textwidth]{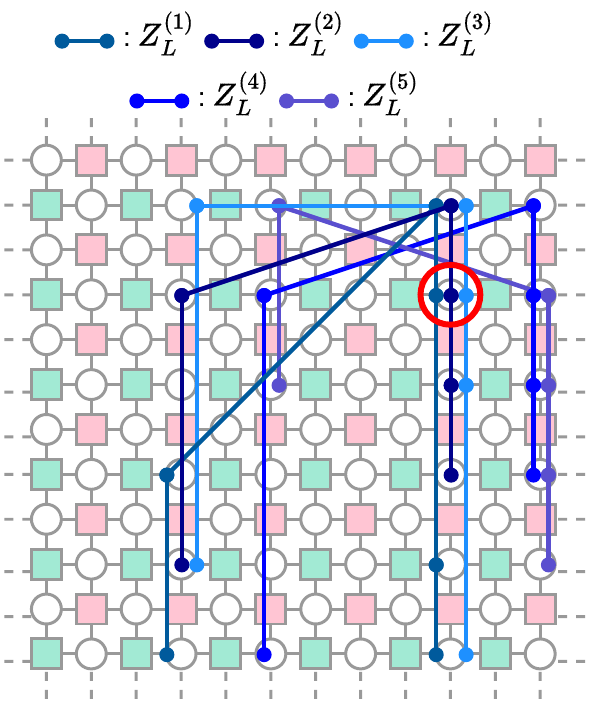}
        \label{fig:logical_bb}
    \end{subfigure}
    \caption{(Left) Logical $Z$ observable support in the $d=5$ surface code. (Right) Supports for a subset of the logical $Z$ observables in the [[72,12,6]] BB code.}
    \label{fig:logical_observables}
    \vspace{-0.2cm}
\end{figure}

\begin{figure*}[h!]
    \centering
    \includegraphics[width=\textwidth]{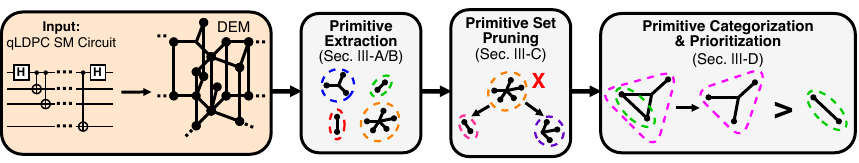}
    \caption{The workflow through which Arqade constructs an irreducible set of predecoding primitives from an input SM circuit.}
    \label{fig:primitive-generation}
\end{figure*}

\subsection{Motivation: Predecoding for qLDPC Codes}
Since BP decoders are typically both slower and more expensive than their surface code counterparts, the problem of how to scale qLDPC-based systems is all the more pressing. Despite their promise for low-cost mitigation of decoder contention, NSM predecoders have only ever been demonstrated for the surface code. This is primarily because general qLDPC codes often feature higher-weight and/or long-range qubit interactions that induce significantly more complex error propagation, breaking the geometric locality implicitly relied upon in prior surface code designs. Building predecoders for general qLDPC codes requires addressing two problems.

\textbf{Problem 1:} It is unclear whether the error sparsity needed for effective predecoding holds under high-complexity error propagation in general qLDPC codes. To answer this, we simulated $10^5$ shots of the BB and color code SM circuits, recording the distribution of maximum-length errors. We define maximum error length as the longest chain of adjacent decoding graph hyperedges triggered per simulation shot.

Fig. \ref{fig:error-length-dist} shows that low-length errors, length-1 in particular, dominate the distribution. At $p=10^{-4}$, length-1 errors represent well over 95\% of maximum-length errors across all code distances. Even at present-day $p=10^{-3}$, length-1 errors represent at least 25\% across all codes, often exceeding 40-80\%. This suggests that error sparsity is a shared property, and indeed highly prevalent, across qLDPC codes. Therefore, effective qLDPC predecoders should provide the same degree of benefit as has been demonstrated for the surface code.

\textbf{Problem 2:} The structure of general qLDPC codes is much more complex and varied than that of the surface code. Fig. \ref{fig:logical_observables} shows that a single error (red circle) in the surface code flips at most one logical observable, while, in the BB code, it can flip three of the shown observables, simultaneously. Furthermore, in the surface code, a given error flips up to 2 syndromes, while a given syndrome is flipped by up to 19 errors; in the [[72,12,6]] code, these dual degeneracies increase significantly to 6 and 42, respectively (not shown in Fig. \ref{fig:logical_observables}). This complexity is difficult for any decoding system, but particularly so for predecoders leveraging only local information. Finally, since the structure of different qLDPC codes is so varied \cite{ErrorCorrectionZoo}, the techniques used to `hand' construct a predecoder for one code are not directly transferrable to another.

Taken together, these challenges render the manual design and hand-optimization techniques used in past predecoders \cite{ravi2023better, alavisamani2024promatch, knapen2026pinball} impractical for general qLDPC codes, necessitating automated, universal methods for predecoder construction. In response, we choose to investigate the automated development of NSM predecoders for dominant length-1 errors in general qLDPC codes, aiming to achieve significant impacts with minimal design complexity.
\section{Automated Predecoder Construction} \label{sec:predecoder-construction}
In this section, we illustrate how the Arqade framework can generate predecoding logic capable of correcting length-1 errors in the decoding graph for any qLDPC code. As shown in Fig. \ref{fig:primitive-generation}, Arqade takes as input the SM circuit for the code and the circuit-level noise model used to describe what types of errors can occur throughout the SM circuit's execution. Then, by raising the level of abstraction from the SM circuit to its corresponding detector error model (DEM) \cite{gidney2021stim}, it uses a series of automated steps to produce a minimal set of predecoding primitives, organized by a two-level hierarchical prioritization scheme, that provide full predecoding coverage for the input qLDPC code.

\subsection{Defining a Predecoding Primitive} \label{sec:preliminaries}
Building on the terminology of Ref. \cite{knapen2026pinball}, we abstract the definition of a predecoder into a collection of predecoding primitives, each of which covers a particular edge of the code's decoding graph. However, unlike Ref. \cite{knapen2026pinball}, we focus on mapping primitives to observables as opposed to explicit data qubits. Within this formalism, a predecoding primitive can be uniquely defined via two sets, $S$ and $O$. $S$ is the set of syndrome bits at the endpoints of the edge on which the predecoding primitive is instantiated, whereas $O$ is the set of logical observables that would be flipped if the primitive assigns a correction to its edge. Each predecoding primitive follows the same, basic update rules. Namely, it first checks if all its syndromes are active (i.e., $s = 1 \; \forall s \in S$). If so, it flips its observables in $O$ and clears its syndromes. Within the code, these observable flips are equivalent to effective data qubit corrections.

\begin{figure}
    \centering
    \begin{subfigure}{0.45\columnwidth}
        \includegraphics[width=\textwidth]{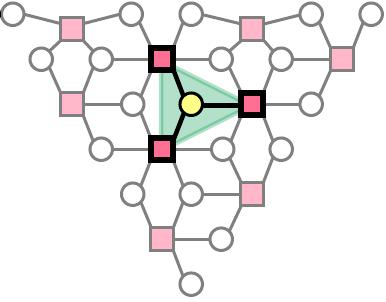}
    \end{subfigure}
    \hspace{0.5cm}
    \begin{subfigure}{0.45\columnwidth}
        \includegraphics[width=\textwidth]{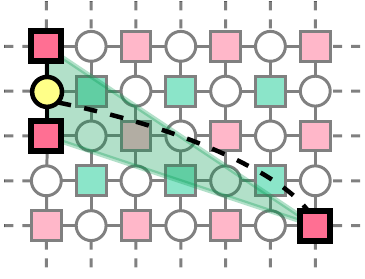}
    \end{subfigure}
    \caption{Examples predecoding primitives in the (left) color code and (right) BB codes. Syndrome sets are bolded, and yellow circles are effective data qubit corrections. The dashed line in the BB code is the relevant long-range connection. For clarity, only $X$ check qubits are shown for the color code.}
    \label{fig:primitives}
\end{figure}

Fig. \ref{fig:primitives} shows example predecoding primitives for the color and BB codes. In both cases, $|S| = 3$, reflecting the presence of hyperedges in the decoding graphs for these codes that preclude the use of matching based decoders like MWPM.

\subsection{Deriving a Complete Primitive Set}
Having defined the structure of a single predecoding primitive, we now discuss how Arqade automatically generates a complete set of predecoding primitives providing full coverage of length-1 errors for a given qLDPC code.

Given the target code's SM circuit as well as the noise model specification, Arqade extracts its corresponding detector error model (DEM) \cite{gidney2021stim}, a representation of the decoding graph providing a straightforward definition of errors in the qLDPC code. Namely, the DEM includes metadata for each error specifying which syndromes it activates, which logical observables it flips, and with what probability the error occurs. 

For each error in the DEM, Arqade records its syndromes and logical observables, mapping them to corresponding indices within \emph{global syndrome and logical observable buffers} shared among all primitives. Arqade builds the sets $S$ and $O$ from these mapped indices to instantiate a new predecoding primitive for the error. 

Using the indices in $S$, the corresponding values in the global syndrome buffer are routed to the inputs of the new predecoding primitive. Conditioned on these syndrome values all being active, Arqade routes the primitive's syndrome updates back to the global syndrome buffer. It also uses the indices in $O$ to route observable updates to the correct locations in the global logical observable buffer. 

If, after executing all predecoding primitives, there remain non-zero bits in the global syndrome buffer, Arqade's corrections did not account for all errors, so it flags the decoding problem as complex (i.e., beyond its localized predecoding capability) and defers the original syndrome data to the second-level decoder.

\subsection{Pruning Strategies} \label{sec:pruning-strategies}
Although the DEM's structure is simplified by error degeneracy, it can still contain many thousands of edges for suitably large and complex codes, yielding a resource-intensive predecoder. Arqade exploits two pruning strategies to reduce the predecoder into a much smaller, irreducible set of primitives.

First, the repetitive application of the SM circuit induces identical per-round structure in the DEM, leading to many predecoding primitives which are copies of each other across different SM rounds. To cover all length-1 errors, the predecoder only needs to process two SM rounds at a time, allowing the same predecoding primitive to be reused for each copy of a DEM edge per two-round block. This reduces the number of primitives by a factor of $d$.

The second pruning strategy filters out \textit{composite primitives} which are effectively a combination of two or more smaller primitives. More formally, given a candidate primitive with syndrome and observable sets $S_C$ and $O_C$, we first find all primitives whose syndrome sets, $S_i$, satisfy $S_i \subset S_C$. Within this group, if there exists a combination of primitives satisfying $S_1 \cup S_2 \cup... \cup S_k = S_C$, and whose combined action on the code's logical observables is equivalent to the candidate ($O_1 \oplus O_2 \oplus ... \oplus O_k = O_C$), then the candidate is a composite primitive and can be removed.

\subsection{Prioritization Scheme} \label{sec:prioritization-scheme}
After generating the minimal set of predecoding primitives, Arqade performs a final step in which it organizes primitives into groups and applies a two-level hierarchical prioritization scheme across and within groups. Primitives assigned higher priority must be executed before those with lower priority. This is essential to implement the predecoder in hardware (Sec. \ref{sec:hardware-arch}).

Arqade first organizes each predecoding primitive into one of a set of four distinct classes: \circled{1} \textit{time-like} primitives, which cover measurement errors on check qubits, \circled{2} \textit{space-like} primitives, which cover errors on one data qubit that activates multiple syndrome values within the same SM round, \circled{3} \textit{spacetime-like} primitives, which are similar, but whose syndrome sets include one or more syndrome values from different SM rounds, and \circled{4} \textit{hook-like} primitives, which cover check qubit errors that propagate to multiple data qubits. We emphasize that assignment of a primitive to any of these definitions is possible using information from the DEM.

To maximize predecoding performance, Arqade then applies the following level-1 priority heuristic: \emph{primitives that belong to classes that cover more common errors should be given higher priority}. We accumulate the probabilities of each DEM error to derive an average probability per error in each class indicating how common errors of that class are, on average.

Tab. \ref{tab:error-class-frequencies} shows these probabilities for a representative sampling of codes under the SI1000 noise model \cite{gidney2021fault}. Time-like errors are consistently most common, followed by space-like, spacetime-like, and hook-like errors. Thus, Arqade assigns highest priority to time-like primitives (signaling they should execute first), second-highest priority to space-like primitives, and so on. We note that, although different noise models may produce different priority orderings, Arqade's specification is flexible to these changes via a reassignment of priorities.

\begin{table}
\centering
\caption{The distribution of error probabilities per primitive class in the SI1000 noise model across representative instances of the surface, color, and BB codes. A higher value indicates that an error within the class is more common, on average.}
\label{tab:error-class-frequencies}
\begin{tabular}{|c|c|c|c|}
\hline
\textbf{\begin{tabular}[c]{@{}c@{}}Predecoding \\Primitive Class\end{tabular}} &
  \textbf{\begin{tabular}[c]{@{}c@{}}Surface Code\\ (d=11)\end{tabular}} &
  \textbf{\begin{tabular}[c]{@{}c@{}}Color Code\\ (d=13)\end{tabular}} &
  \textbf{\begin{tabular}[c]{@{}c@{}}BB Code\\ {[[144,12,12]]}\end{tabular}} \\ \hline
Time-like &
  \begin{tabular}[c]{@{}c@{}}$1.34 \cdot 10^{-4}$\end{tabular} &
  \begin{tabular}[c]{@{}c@{}}$4.19 \cdot 10^{-4}$\end{tabular} &
  \begin{tabular}[c]{@{}c@{}}$4.35 \cdot 10^{-4}$\end{tabular} \\ \hline
Space-like &
  \begin{tabular}[c]{@{}c@{}}$4.43 \cdot 10^{-5}$\end{tabular} &
  \begin{tabular}[c]{@{}c@{}}$1.54 \cdot 10^{-4}$\end{tabular} &
  \begin{tabular}[c]{@{}c@{}}$1.16 \cdot 10^{-4}$\end{tabular} \\ \hline
Spacetime-like &
  \begin{tabular}[c]{@{}c@{}}$1.33 \cdot 10^{-5}$\end{tabular} &
  \begin{tabular}[c]{@{}c@{}}$2.83 \cdot 10^{-5}$\end{tabular} &
  \begin{tabular}[c]{@{}c@{}}$3.81 \cdot 10^{-5}$\end{tabular} \\ \hline
Hook-like &
  \begin{tabular}[c]{@{}c@{}}$8.99 \cdot 10^{-6}$\end{tabular} &
  \begin{tabular}[c]{@{}c@{}}$2.95 \cdot 10^{-5}$\end{tabular} &
  \begin{tabular}[c]{@{}c@{}}$2.87 \cdot 10^{-5}$\end{tabular} \\ \hline
\end{tabular}
\end{table}

Ordering of predecoding primitives is further complicated by \textit{subset primitives}, or primitives whose syndrome sets are proper subsets of another. Even after applying composite primitive pruning (Sec. \ref{sec:pruning-strategies}), subset primitives may persist within and across each error class. Consider two primitives $P_i$ and $P_j$, derived from DEM error edges $E_i$ and $E_j$, with corresponding syndrome sets $S_i$ and $S_j$, and suppose $S_i \subset S_j$.

Regardless of whether $E_i$ or $E_j$ occurs, all syndromes in $S_i$ will be active. If $P_i$ is ordered first by priority ($P_i \rightarrow P_j$), $P_i$ will always assign a correction and clear the syndromes in $S_i$. Hence, if $E_j$ had occurred, $P_j$ would not trigger, since it would only see the residual syndromes $S_j \setminus S_i$. Such a scenario can severely harm predecoder coverage and accuracy.

Critically, the reverse ordering ($P_j \rightarrow P_i$), guarantees correctness. If $E_j$ occurs, $P_j$ can correct for it, and if $E_i$ occurs, $P_j$ will not be triggered, enabling $P_i$ to correct for it. To ensure this safe ordering, Arqade must apply an additional level-2 prioritization to assign subset primitives lower priorities than their supersets. If both primitives exist within the same class, Arqade assigns a lower within-class priority to the subset primitive. Otherwise, if they exist across different classes, and the subset's class has higher level-1 priority, we move it to a newly created subset primitive class which has lowest priority.

\subsection{Benefits of Design Abstraction}
Our choice to move away from physical circuits and qubits and towards higher-level constructs like the code's DEM and logical observables has several key benefits. First, since errors in the DEM are distinguished only by the sets of activated syndromes and flipped logical observables, Arqade need not distinguish between different circuit-level faults whose effects on the underlying code are equivalent (so-called \textit{degenerate errors}). Instead, only one predecoding primitive is needed to correct for any such degenerate error. In the [[72,12,6]] code, this translates to maximum (average) savings of $44\times$ ($17.36\times$) in the number of predecoding primitives per degenerate set.

Moreover, since the DEM exists at a level of abstraction unifying all qLDPC codes, it decouples the predecoder architectural design process from the underlying target QEC code. Namely, defining predecoding primitives only from information available in the DEM enables them to serve as universal building blocks upon which the predecoder for any qLDPC code can be built. Similarly, categorizing these primitives from information available in the DEM enables the same primitive prioritization scheme to be reused across all qLDPC codes.
\begin{figure*}[h!]
    \centering
    \includegraphics[width=\textwidth]{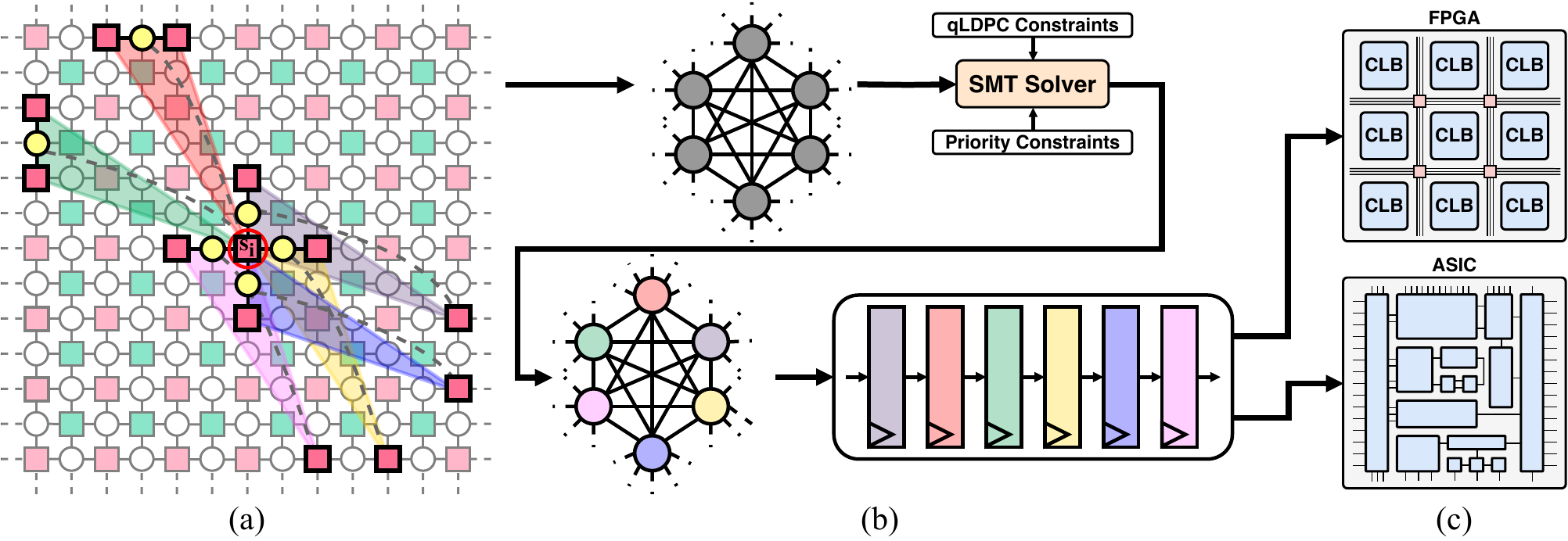}
    \caption{End-to-end flow of Arqade's hardware architecture generation, beginning with (a) a set of conflicting predecoding primitives, indicated by colored triangles, (b) the construction of the conflict graph and application of qLDPC-inspired graph coloring to schedule primitives into a pipeline, and (c) the flexible synthesis of the pipeline onto FPGA or ASIC hardware.}
    \label{fig:conflict-graph}
\end{figure*}

\section{Hardware Architecture} \label{sec:hardware-arch}
With the full predecoding logic generated, we now discuss the second phase of Arqade's automated workflow: mapping this logic into an efficient, real-time hardware architecture. 

From a dataflow perspective, each predecoding primitive executes a read-modify-write (RMW) operation on the bits in the global syndrome buffer. Thus, if the syndrome sets for two predecoding primitives intersect on any bits, there exists a data hazard between them; we term such primitives as \textit{conflicting}. Fig. \ref{fig:conflict-graph}a shows an example of six BB code primitives conflicting on a common syndrome, $s_i$, circled in red. To ensure all primitives maintain a consistent view of the global syndrome buffer, each RMW operation must be atomic, and therefore, any conflicting primitives must be executed sequentially. 

An effective way to manage this atomicity in hardware is by organizing conflicting predecoding primitives into separate pipeline stages. This allows non-conflicting primitives within the same pipeline stage to execute in parallel. Arqade must account for multiple design objectives when constructing these stages. First, minimizing the depth of the pipeline ensures maximal concurrency within each pipeline stage and reduces the predecoder's end-to-end latency, supporting real-time decoding. Second, to maximize predecoding accuracy, it must maintain an ordering between primitives that respects the prioritization scheme introduced in Sec. \ref{sec:prioritization-scheme}.

Simultaneously balancing these two design objectives for arbitrary qLDPC codes poses several significant challenges. Most fundamentally, manual pipeline optimization techniques for one code do not generalize to other codes. Separate from lacking generalizability, codes with higher-weight stabilizer checks generally produce primitives with larger syndrome sets. Intuitively, such primitives have a higher probability of conflicting with a larger number of other primitives, rendering manual searches for conflict-free arrangements infeasible.

\subsection{Pipeline Construction via Graph Coloring}
 Given these challenges, a systematic, automated strategy is needed to obtain efficient pipelines. To this end, we leverage the well-established link between conflict resolution problems and graph coloring algorithms \cite{leighton1979graph}. We first translate predecoder pipeline construction into a graph coloring problem.

To describe conflicts between predecoding primitives, we use a data structure we call the \textit{conflict graph} (Fig. \ref{fig:conflict-graph}b). In the conflict graph, nodes represent primitives, and an edge connects two nodes if the corresponding primitives conflict. Under this formalism, assigning a color to a node in the conflict graph corresponds to assigning a primitive to a particular pipeline stage, and searching for the minimum number of pipeline stages becomes equivalent to finding the conflict graph's chromatic number, $\chi$ \cite{harary1971graph}.

Finding $\chi$ is NP-hard in general, so performing a search over the conflict graph of \textit{all} predecoding primitives quickly becomes intractable as the size of the code increases. To simplify this problem, Arqade constructs conflict graphs and colors them separately for each primitive class. We maintain compliance with Arqade's prioritization scheme by incorporating level-2 priorities into the coloring procedure for each class and by aggregating generated stages into the final pipeline in order of level-1 priority. However, the conflict graph for a single class may still be too complex to color optimally via brute force search. Thus, we devise a novel graph coloring strategy exploiting properties specific to qLDPC codes to obtain near-optimal results within reasonable runtime.

\subsection{qLDPC-Informed Graph Coloring} \label{sec:smt-solver}
Underpinning our qLDPC-informed graph coloring strategy is the observation that the qLDPC constraint, namely each qubit in the code interacts with a small fixed number of other qubits, induces useful structure in the conflict graph that can be exploited to greatly reduce the graph coloring search space.

Consider again the syndrome qubit $s_i$ shown in Fig. \ref{fig:conflict-graph}a. Due to the BB code's weight-6 stabilizers, $s_i$ interacts with six data qubits (highlighted yellow). As a result, $s_i$ is an element of six distinct space-like primitives' syndrome sets (colored triangles), each covering an error on one of those data qubits, and all six of these primitives mutually conflict with each other. From the perspective of the space-like conflict graph, this group of conflicting primitives forms a clique (Fig. \ref{fig:conflict-graph}b). This scenario repeats for every check qubit in the code, so the conflict graph manifests as a series of interconnected cliques.

Cliques are a particularly useful structure in the context of graph coloring for two reasons. First, since all nodes in a clique must have mutually distinct colors, they impose a stronger constraint over an entire set of nodes rather than the simpler constraint over pairwise neighbors. Second, the largest clique in a graph, $\omega(G)$, sets a lower bound on the graph's chromatic number: $\omega (G) \leq \chi$. Thus, finding the largest clique in the conflict graph bounds from below our search for a minimum-depth predecoding pipeline for a given qLDPC code.

We use these facts to efficiently search for an $\omega(G)$-coloring of the conflict graph using Z3's \cite{de2008z3} Satisfiable Modulo Theories (SMT) solver. The following two constraints define the $\omega(G)$-coloring problem:
\begin{align*}
    \forall \text{ nodes } u \in G &: 1 \leq color(u) \leq \omega(G) \\
    \forall \text{ edges } (u, v) \in G &: color(u) \neq color(v) 
\end{align*}

\noindent
The first constraint restricts the solver's search to using the lower bound set by the conflict graph's largest clique, and the second is the basic graph coloring constraint. We use networkx's \cite{networkx} \texttt{max\_weight\_clique()} function to quickly find the largest clique in the conflict graph.

To preserve Arqade's within-class prioritization (Sec. \ref{sec:prioritization-scheme}), we add constraints to ensure that lower-priority predecoding primitives land in later pipeline stages.
\begin{align*}
    priority(u) \lessgtr priority(v) &\implies color(u) \gtrless color(v) 
\end{align*}

\noindent
Finally, we find additional cliques in the conflict graph to add stronger clique-coloring constraints to the solver. However, complex error propagation in high-weight qLDPC codes increases connectivity in their conflict graphs, yielding exponentially many cliques whose constraints overwhelm solver memory if enumerated exhaustively. Instead, we search for a smaller set $C_k$ of the $k=16$ largest cliques in the conflict graph.
\begin{align*}
    \forall C \in C_{k} &: \bigwedge_{u,v \in C} color(u) \neq color(v)
\end{align*}

\noindent
We find $C_k$ using a greedy algorithm that progressively grows a clique using the adjacent node sharing the most neighbors with that clique. In practice, this restricted search is effective in speeding up the solver, since constraints derived from the largest cliques prune the search space by the greatest amount.

Despite its promise, the SMT solver may not converge to an $\omega(G)$-coloring of the conflict graph. First, the maximal clique size provides only a lower bound for the $\chi$, so an $\omega(G)$-coloring of the graph may not exist. In this scenario, Arqade employs a configurable retry procedure where, on the $r$th retry, it allows the solver to use $\omega(G)+r$ colors in its solution. Second, the clique constraints and symmetry breaking may not reduce the search space enough to enable convergence within a reasonable amount of time. For this case, Arqade utilizes a configurable timeout, set to 10 minutes in this work.

If the solver fails to converge within that timeout, Arqade falls back to a set of generic, greedy graph coloring algorithms \cite{kosowski2004classical, matula1983smallest, deo2006discrete}, selecting the one that uses the fewest colors. As these generic graph coloring algorithms employ their own heuristics (e.g., node degree) to assign node priorities, they may violate Arqade's prioritization and should be avoided wherever possible to maintain predecoding performance. This can be helped by increasing $k$, $r$, or the timeout for the solver. Since pipeline construction is a one-time overhead per code, a much larger timeout than 10 minutes can be tolerated to give the best chance of finding an optimal solution. In practice, we find that the solver converges for almost all conflict graphs within seconds, struggling only with BB and GB codes which feature long-range interactions and high connectivity.

\subsection{Additional Pipeline Truncation} \label{sec:pipeline-stage-removal}
In Tab. \ref{tab:pipeline-depths}, we show the maximum pipeline depth achieved by Arqade for a variety of codes. In general, codes with higher-weight stabilizers require much deeper pipelines. This aligns with the intuition from Sec. \ref{sec:smt-solver}: the more data qubits a check qubit interacts with, the more primitives will conflict on that check qubit's syndrome. This increase is very manageable owing to the lightweight implementation of predecoding primitives and relaxed parallel window decoding latency budgets. 

Alternatively, if end-to-end latency is a strict constraint, deeper pipelines will necessitate higher operating frequency which increases power consumption. Here, Arqade's prioritization scheme allows us to avoid frequency and power increases by removing stages from the pipeline. Since primitive classes are mapped to pipeline stages in order from highest to lowest priority, if stages are removed from the tail of the pipeline, the priority ordering between stages does not change. Consequently, while the predecoder's coverage will decrease, as some errors will not be checked by a primitive, its logical error rate should not be affected. This creates a direct tradeoff between coverage and hardware complexity (see Sec. \ref{sec:coverage-vs-complexity}).

\begin{table}
\centering
\caption{Depths of the pipelines constructed by Arqade for a variety of qLDPC codes.}
\label{tab:pipeline-depths}
\begin{tabular}{|c|c|c|c|}
\hline
\begin{tabular}[c]{@{}c@{}}\textbf{Code Family}\end{tabular} &
  \textbf{Instance} &
  \begin{tabular}[c]{@{}c@{}}\textbf{Stabilizer Weight}\end{tabular} &
  \begin{tabular}[c]{@{}c@{}}\textbf{Pipeline Depth}\end{tabular} \\ \hline
\begin{tabular}[c]{@{}c@{}}Surface Code\end{tabular}              & $d=3-15$               & 4 & 9  \\ \hline
\begin{tabular}[c]{@{}c@{}}Color Code\end{tabular}                & $d=3-15$               & 6 & 22-30 \\ \hline
\multirow{3}{*}{\begin{tabular}[c]{@{}c@{}}BB Codes\end{tabular}} & $[[72,12,6]]$  & 6 & 19 \\ \cline{2-4} 
                          & $[[108,8,10]]$ & 6 & 24 \\ \cline{2-4} 
                          & $[[144,12,12]]$      & 6 & 21 \\ \hline
\multirow{3}{*}{\begin{tabular}[c]{@{}c@{}}GB Codes\\ \cite{bravyi2024high, lin2024quantum}\end{tabular}} & $[[90,8,10]]$ & 6 & 30 \\ \cline{2-4} & $[[72,8,10]]$  & 8 & 80 \\ \cline{2-4} 
                          
                          & $[[96,10,12]]$      & 8 & 69 \\ \hline

\multicolumn{1}{|c|}{\multirow{3}{*}{{\begin{tabular}[c]{@{}c@{}}RQT Codes\\ \cite{radebold2025explicit, viszlai2026prophunt}\end{tabular}}}} & $[[54,12,4]]$    & 6                                 & 27                                                   \\ \cline{2-4} 
\multicolumn{1}{|l|}{}    & $[[108,18,4]]$ & 6 & 30 \\ \cline{2-4} 
\multicolumn{1}{|l|}{}    & $[[60,2,6]]$   & 4 & 11 \\ \hline
\end{tabular}
\end{table}

\begin{figure*}[ht!]
    \centering
    \includegraphics[width=\textwidth]{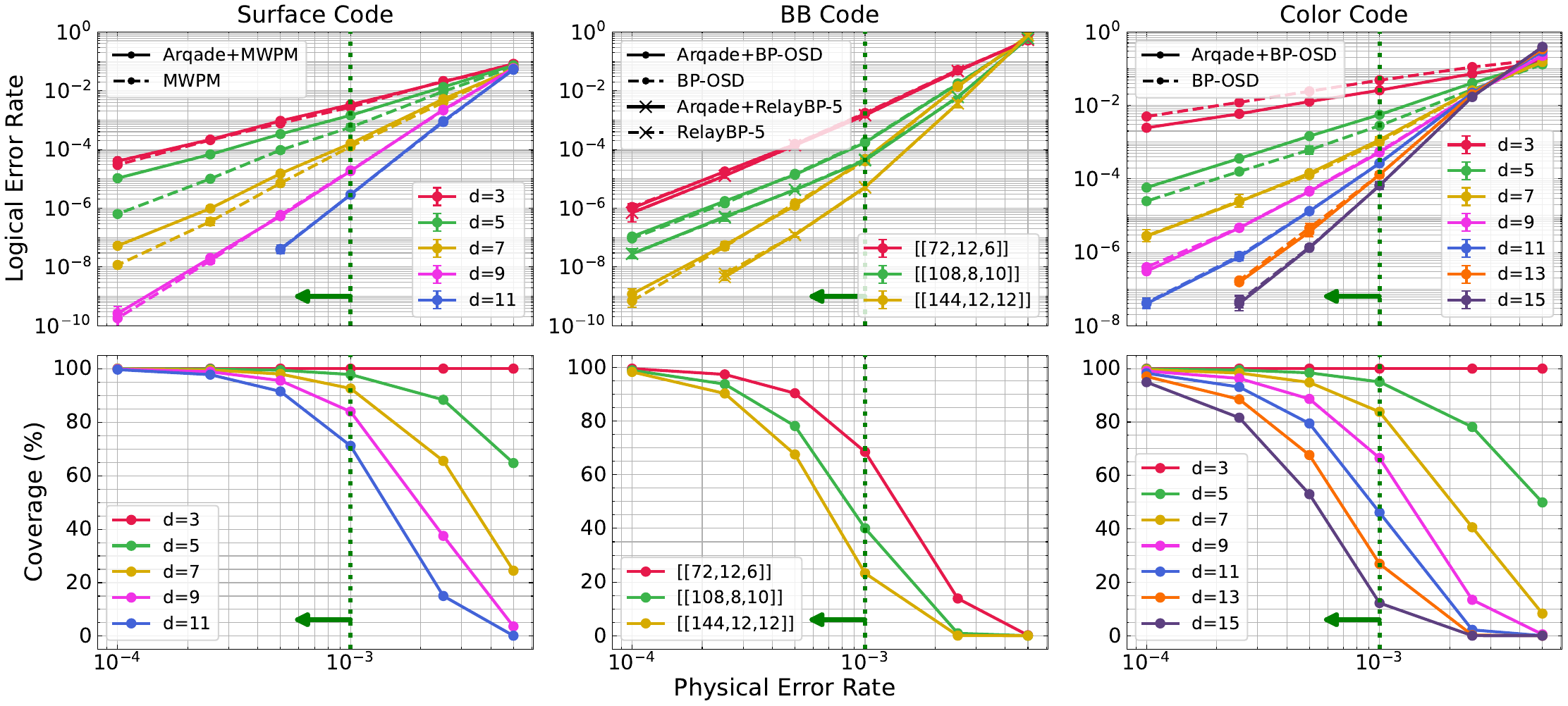}
    \caption{Arqade's performance over a range of codes, including (top) LER comparisons between a system using Arqade as a predecoder (solid lines) and a system using only a decoder (dashed lines), and (bottom) Arqade's predecoding coverage. Physical error rates to the left of the dotted line correspond to the QEC-relevant regime.}
    \vspace{-0.5cm}
    \label{fig:ler_coverage}
\end{figure*}

\subsection{Generalizable and Adaptable Synthesis}
Arqade's pipelined architecture allows for flexible implementations tailored to different design scenarios (Fig. \ref{fig:conflict-graph}c). For example, when architecting for reconfigurable quantum computing systems like neutral atom or trapped ion qubits, designers may want to explore different codes and SM circuits, given that the state-of-the-art is rapidly evolving. For such systems, it is straightforward to synthesize the predecoder onto a room-temperature FPGA. Arqade can then generate predecoders for new codes and circuits to be programmed onto the same FPGA hardware, significantly reducing design iteration costs. While low latencies can be challenging for FPGA hardware with lower maximum frequency ($\sim$100s of MHz), pipeline stage removal may mitigate this.

Alternatively, resource-constrained systems like superconducting qubits with static, limited connectivity can reasonably support a smaller space of qLDPC codes. Here, system architects may prefer aggressive optimization at each layer of the stack. In this case, the predecoder can be synthesized to a highly optimized ASIC compatible with operation inside the dilution refrigerator. Again, for this design scenario, fine-grained optimization via pipeline stage removal may simplify cryogenic ASIC implementation, since lower operating frequency, fewer pipeline registers, and less switching activity all reduce power consumption overheads.
\section{Evaluation} \label{sec:evaluation}
We demonstrate Arqade's generalizability and performance across the eleven qLDPC codes listed in Tab. \ref{tab:pipeline-depths}, each of which has diverse properties. These include planar codes with open boundary conditions (surface and color codes), high-rate codes with long-range connections (BB, GB, and RQT codes), and different stabilizer weights (from 4 to 8). 

While Arqade is compatible with any second-level decoder, we use the state-of-the-art MWPM-based Pymatching \cite{higgott2025sparse} for the surface code. In most other cases, we use BP-OSD \cite{higgot2025stimbposd} due to its high performance over the widest range of qLDPC codes. The only exceptions are (1) the BB codes, for which we also compare with RelayBP-5 \cite{muller2025improved}, and (2) the GB codes, for which we exclusively use RelayBP-5, since it achieves much higher accuracy than BP-OSD. For BP-OSD, we set the number of BP iterations assuming a 20 ns iteration time \cite{muller2025improved} and a total BP latency budget of $d$ $\mu s$ for a distance-$d$ qLDPC code. For RelayBP-5, we use its default parameters.

We perform memory experiment simulations in Stim \cite{gidney2021stim} which include $d$ rounds of syndrome measurement followed by a final measurement of all data qubits. For the surface code, we use the optimal N-Z SM circuit \cite{tomita2014low}, for the BB codes, we use IBM's SM circuit \cite{bravyi2024high}, and for the color code, we use the off-the-hook SM circuit \cite{kishony2026color}. For all other codes, we use the qLDPC library \cite{perlin2023qldpc} to generate coloration SM circuits \cite{delfosse2023constant}. Our simulations use the hardware-realistic SI1000 noise model \cite{gidney2021fault} which accounts for errors due to single- and two-qubit gates, measurement, reset, and idling. Each error channel is parameterized by a base physical error rate, $p$, which we scale in the range $[10^{-4}, 5 \cdot 10^{-3}]$. This includes error rates in present-day hardware \cite{acharya2024quantum} and projected for future devices.

To characterize Arqade's hardware efficiency, we perform post-synthesis analysis for two design scenarios: a room temperature AMD Zynq UltraScale+\texttrademark{} ZCU102 FPGA \cite{amd2023zcu}, and a cryogenic ASIC implemented in a 4K-characterized, 22 nm FDSOI technology node. We use FPGA LUT and register utilization as a proxy for hardware overheads, while for the ASIC design, we conduct standard PPA analysis. We restrict our hardware evaluations to surface, color, and BB codes.

\begin{table}[t]
\centering
\caption{Coverage and logical error rate comparisons for additional qLDPC codes. Each cell reports the value at $p=10^{-4}$, with the value at $p=5\cdot10^{-4}$ in parentheses.}
\label{tab:additional-codes}
\begin{tabular}{|c|ccc|}
\hline
\textbf{Instance} &
  \multicolumn{1}{c|}{\textbf{Coverage}} &
  \multicolumn{1}{c|}{\textbf{Arqade+L2 LER}} &
  \textbf{L2-Only LER} \\ \hline
{[}{[}72,8,10{]}{]} &
  \multicolumn{1}{c|}{96.2 (59.2)} &
  \multicolumn{1}{c|}{1.3E-6 (2.1E-4)} &
  1.3E-6 (2.1E-4) \\ \hline
{[}{[}90,8,10{]}{]} &
  \multicolumn{1}{c|}{97.9 (74.8)} &
  \multicolumn{1}{c|}{7.2E-8 (1.6E-5)} &
  7.2E-8 (1.6E-5) \\ \hline
{[}{[}96,10,12{]}{]} &
  \multicolumn{1}{c|}{95.8 (48.2)} &
  \multicolumn{1}{c|}{5.2E-8 (2.7E-5)} &
  5.2E-8 (2.7E-5) \\ \hline
{[}{[}54,12,4{]}{]} &
  \multicolumn{1}{c|}{98.8 (89.6)} &
  \multicolumn{1}{c|}{6.8E-4 (1.1E-2)} &
  6.7E-4 (1.1E-2) \\ \hline
{[}{[}108,18,4{]}{]} &
  \multicolumn{1}{c|}{97.9 (79.9)} &
  \multicolumn{1}{c|}{3.3E-4 (8.1E-3)} &
  2.9E-4 (7.5E-3) \\ \hline
{[}{[}60,2,6{]}{]} &
  \multicolumn{1}{c|}{99.8 (94.9)} &
  \multicolumn{1}{c|}{6.9E-6 (3.2E-4)} &
  7.1E-6 (3.2E-4) \\ \hline
\end{tabular}
\end{table}

\subsection{Predecoder Performance}
In Fig. \ref{fig:ler_coverage}, we show Arqade's predecoding performance for the surface, BB, and color codes. In the top row, we compare the logical error rates (LERs) of two systems: one with a decoding hierarchy (the \texttt{Arqade+L2} configuration), and the other with only a second-level decoder (the \texttt{L2-only} configuration). In the bottom row, we show Arqade's coverage, or the percentage of the decoding workload it handles without having to defer to the second-level decoder. Similar statistics are reported over a wider range of code instances in Tab. \ref{tab:additional-codes}.

\subsubsection{Logical Error Rate}
Across all codes and decoders, the \texttt{Arqade+L2} configuration achieves LER on par with the \texttt{L2-only} configuration. The only exceptions are at $d=5,7$ in the surface code and $d=5$ in the color code, but the increases are only marginal. In both cases, low-distance codes with open boundary conditions challenge Arqade's complex detection logic, as a higher fraction of leftover syndromes can be matched to the code boundaries. At higher code distances, this fraction lowers, and \texttt{Arqade+L2} recovers parity with \texttt{L2-only}. Overall, these results demonstrate Arqade's ability to maintain state-of-the-art QEC performance across codes.

\subsubsection{Coverage}
Arqade consistently and drastically reduces the percentage of the decoding workload processed by the second-level decoder. Across the surface, BB, and color codes, within the regime relevant for QEC deployment ($p < 10^{-3}$), Arqade achieves a maximum (minimum) coverage of 99.98\% (91.48\%), 99.57\% (67.57\%), and 99.87\% (52.98\%), respectively; at $p=10^{-4}$, Arqade always achieves $\geq 95\%$ coverage across all codes. As $p$ increases, errors become too dense to resolve locally, but Arqade gracefully balances aggressive coverage with high-accuracy decoding, gradually decreasing coverage to maintain LER on par with \texttt{L2-only}.

Arqade's coverage translates to significant benefits at multiple layers in the QCI. First, Arqade reduces second-level decoder utilization at $p=10^{-4}$ by up to 4,090$\times$, $232 \times$, and $750\times$ for the surface, BB, and color codes, respectively. Even at present-day $p=10^{-3}$, Arqade achieves up to $45.58\times$, $3.18\times$, and $20.15\times$ reduction. Keeping the number of decoders constant, this eases contention for shared decoding resources by several orders of magnitude. Equivalently, more logical qubits can be assigned per decoder, requiring orders of magnitude fewer decoders. For systems with cryogenic qubits, deploying Arqade on a cryogenic ASIC (which we show is feasible in Sec. \ref{sec:hw-implementation}) also reduces the required syndrome transmission bandwidth and power.

\subsubsection{Comparison with Prior Work}
\begin{figure}
    \centering
    \includegraphics[width=\columnwidth]{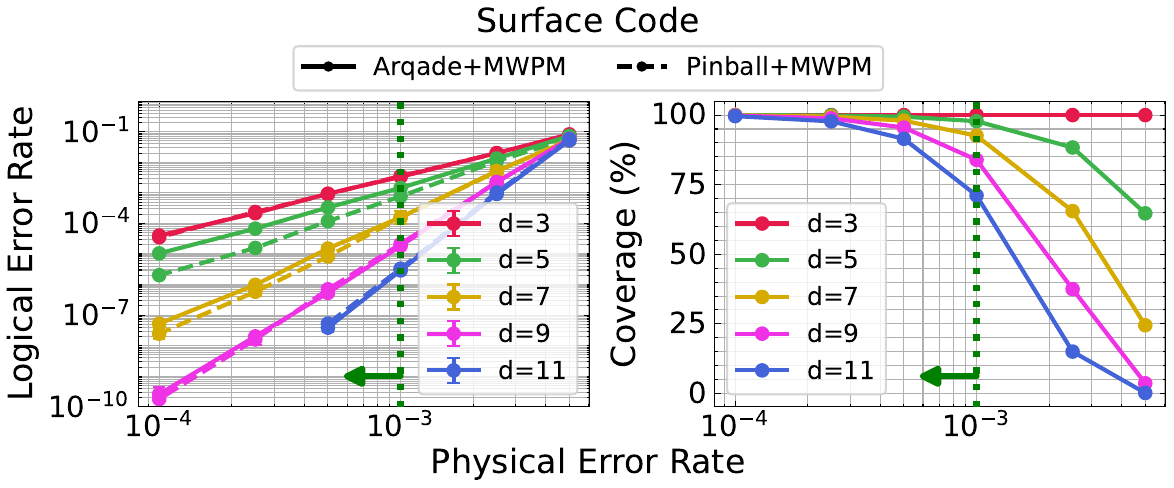}
    \caption{Comparison between Arqade and Pinball \cite{knapen2026pinball}.}
    \label{fig:arqade-vs-pinball}
\end{figure}

In Fig. \ref{fig:arqade-vs-pinball}, we compare Arqade's performance to Pinball \cite{knapen2026pinball}, the state-of-the-art, hand-optimized NSM predecoder for the surface code. Across all code distances, Arqade achieves identical coverage to Pinball, and apart from $d=5,7$, it also achieves identical LER. When analyzing scenarios where Arqade's decoding fails and Pinball's succeeds, we found they were due to code-specific optimizations in its ordering of space-like primitives at the code boundaries. Again, since the impacts of boundary effects diminish as $d$ increases, the LER gap between the predecoders disappears beyond $d=7$. 

Although Arqade's optimizations are necessarily limited to code-agnostic strategies, these results prove their efficacy. Thus, Arqade achieves near-optimal performance while avoiding labor-intensive, manual optimization techniques that lack generalizability to other codes. In the long term, if system architects settle on specific codes and/or SM circuits for which fine-grained optimization is well-motivated, Arqade can still serve as an effective baseline on which to improve.

\subsection{Additional Benefits for L2 Decoding}
As established in Sec. \ref{sec:decoding}, slowdown due to OSD post-processing in BP-OSD or long, sequential BP legs in RelayBP worsen resource contention. Thus, any additional reduction in these level-2 bottlenecks provided by Arqade is crucial for minimizing resource costs in the QCI. 

To explore this, we perform parallel simulations of Arqade, BP-OSD, and RelayBP-5. For each shot, we record whether Arqade provides a full decoding solution or not. Simultaneously, for BP-OSD, we record when its BP step fails to converge, and OSD is invoked. To give BP the best chance to converge, we generously allocate $10\times$ the number of BP iterations normally available within a $d$ $\mu s$ latency budget. For RelayBP-5, we record when it requires extra ($>$5) relay legs to converge, thus occupying the given BP decoder for a longer period of time. Given these, we define two metrics: \textit{OSD reduction} for BP-OSD, the percentage of shots where BP fails to converge but Arqade predecodes, and \textit{relay leg reduction} for RelayBP-5, the percent difference between the number of extra relay legs RelayBP-5 needs with and without Arqade.

\begin{figure}
    \centering
    \begin{subfigure}{0.85\columnwidth}
        \includegraphics[width=\columnwidth]{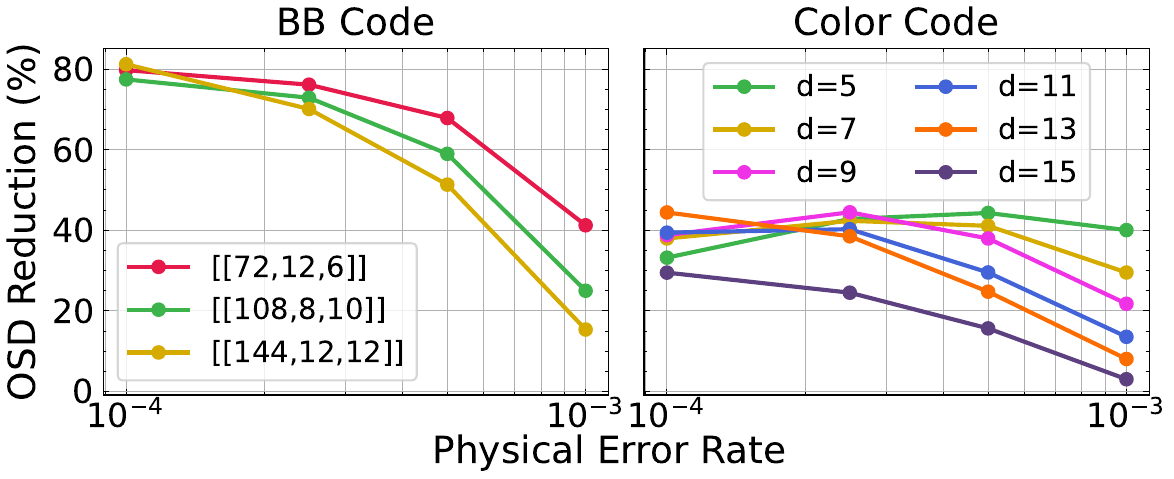}
        \label{fig:osd-reduction}
    \end{subfigure}
    \begin{subfigure}{0.75\columnwidth}
        \includegraphics[width=\columnwidth]{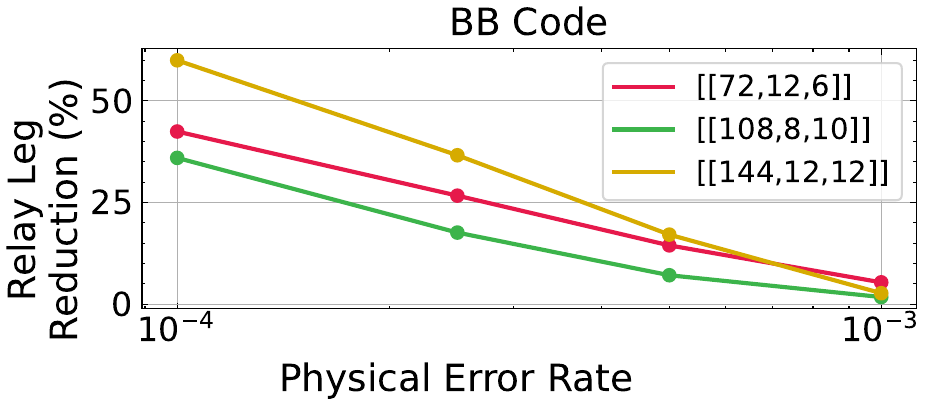}
        \label{fig:relay-reduction}
    \end{subfigure}
    \vspace{-0.5cm}
    \caption{Reduction in (top) OSD utilization and (bottom) extra relay legs needed for convergence when using Arqade.}
    \label{fig:l2-benefits}
\end{figure}

Fig. \ref{fig:l2-benefits} shows that significant OSD and relay leg reductions are indeed possible across qLDPC codes. Arqade achieves OSD reduction up to 81.19\% (44.42\%) across the BB (color) codes, and relay leg reduction up to 59.96\% for the BB codes. Not only do both benefits reduce the average decoding latency per logical qubit, they also open up the possibility to further share OSD instances among multiple BP instances, or BP instances among more logical qubits, both of which can further bring down compute costs for decoding.

We believe these reductions are possible due to how the two decoding levels handle short, localized cycles in qLDPC codes. BP-based decoders struggle with these cycles, called trapping sets, since they cannot confidently decide between two or more possible errors. Hence, final decisions must be deferred to additional compute to maintain low LER. By contrast, since Arqade can always fall back to the global decoder, it can more aggressively commit to one local error over another, and in practice, it reaches the correct solution in a non-trivial percentage of such scenarios.

As physical error rates increase, Arqade's coverage decreases, causing both OSD and relay leg reduction to decrease. In some color code instances (e.g., $d=5$), Arqade's coverage at higher error rates decreases more slowly than BP's convergence failures, causing slight increases in OSD reduction.

\subsection{Hardware Implementation} \label{sec:hw-implementation}
\begin{table}
\centering
\caption{Arqade's FPGA Lookup Table (LUT) and register utilization for large-instance codes.}
\label{tab:fpga-results}
\begin{tabular}{|c|c|c|}
\hline
\textbf{Code Instance}              & \textbf{LUT Utilization} & \textbf{Register Utilization} \\ \hline
{[}{[}225,1,15{]}{]} Surface Code & 0.29\%                   & 0.50\%                           \\ \hline
{[}{[}169,1,15{]}{]} Color Code   & 0.53\%                   & 0.91\%                        \\ \hline
{[}{[}144,12,12{]}{]} BB Code     & 0.66\%                   & 0.57\%                         \\ \hline
\end{tabular}
\end{table}

\begin{table}
\centering
\caption{Arqade's power consumption and area utilization when implemented as a cryogenic ASIC.}
\label{tab:asic-results}
\begin{tabular}{|c|cc|c|}
\hline
\multirow{2}{*}{\textbf{Code Instance}} &
  \multicolumn{2}{c|}{\textbf{\begin{tabular}[c]{@{}c@{}}Average Peak Power \\ @ $p=10^{-3}$\end{tabular}}} &
  \multirow{2}{*}{\textbf{Area Utilization}} \\ \cline{2-3}
                                    & \multicolumn{1}{c|}{100 ns}  & 1 us    &                             \\ \hline
{[}{[}225,1,15{]}{]} Surface Code & \multicolumn{1}{c|}{0.12 mW} & 12 $\mu W$   & 0.0038 $mm^2$ \\ \hline
{[}{[}169,1,15{]}{]} Color Code   & \multicolumn{1}{c|}{0.26 mW} & 26.6 $\mu W$ & 0.0068 $mm^2$ \\ \hline
{[}{[}144,12,12{]}{]} BB Code     & \multicolumn{1}{c|}{0.17 mW} & 16.9 $\mu W$   & 0.0049 $mm^2$  \\ \hline
\end{tabular}
\end{table}

Tab. \ref{tab:fpga-results} shows Arqade's peak LUT and register utilization on the ZCU102 FPGA for each of the $d=15$ surface, $d=15$ color, and $[144,12,12]]$ BB codes. In all cases, Arqade consumes $<1\%$ of both LUT and register resources, meaning a single FPGA board can always accommodate $\sim$100 Arqade instances, regardless of the chosen qLDPC code. For BB codes, Arqade supports decoding for $\sim$1,800 logical qubits per FPGA. Alternatively, Arqade's low footprint allows for co-location with other qubit control and readout peripherals on the same FPGA to realize tight, closed-loop integration.

For Arqade's cryogenic ASIC implementation, we consider two end-to-end pipeline latencies: an aggressive 100 ns and a relaxed 1 $\mu$s. Depending on the pipeline depth, we perform PPA analysis with the corresponding frequency needed to meet these latency targets. Tab. \ref{tab:asic-results} shows the results of this analysis at $p=10^{-3}$. Under the 100 ns latency budget, all predecoder instances operate below 0.3 mW average peak power, whereas for the 1 $\mu$s latency budget, this drops to below 30 $\mu$W. We note that power consumption will drop even lower as physical error rates decrease due to lower hardware switching activity. Arqade's power efficiency is supplemented in all cases by low area footprint, enabling cryogenic operation to scale well into the FTQC regime. For example, given a 1.5 W power budget at 4 K \cite{krinner2019engineering}, Arqade supports $\sim$50,000-500,000 logical qubits in the $[[144,12,12]]$ BB code depending on its target latency.

\subsection{Removing Pipeline Stages} \label{sec:coverage-vs-complexity}

\begin{figure}
    \centering
    \includegraphics[width=\columnwidth]{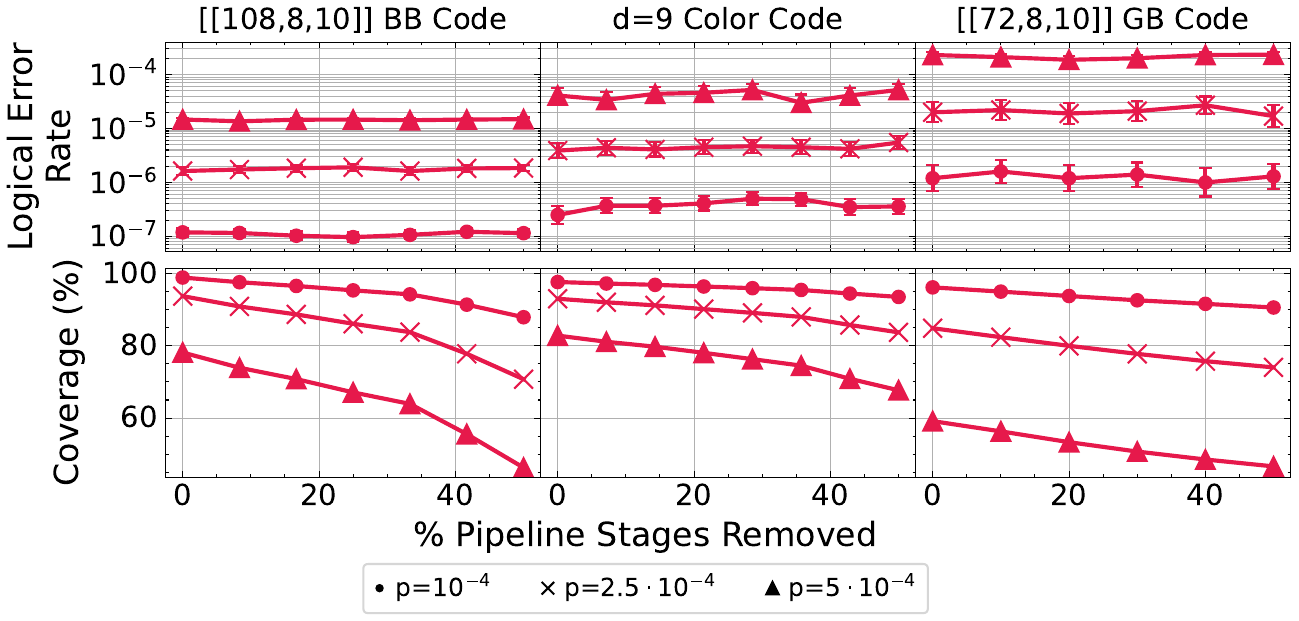}
    \caption{Arqade's LER and coverage as stages are successively removed from the tail of its pipeline.}
    \label{fig:stages_removed}
\end{figure}

Sec. \ref{sec:pipeline-stage-removal} discussed the trade off between power/area and coverage by removing stages from Arqade's pipeline. We explore this tradeoff for color, BB, and GB codes, all of which feature deep pipelines (Tab. \ref{tab:pipeline-depths}) and high initial coverage.

Fig. \ref{fig:stages_removed} shows the effects of removing up to 50\% of pipeline stages across multiple code distances and physical error rates. As expected, removing stages affects Arqade's coverage, but not its LER. At low physical error rates, the tradeoff improves, reflected by shallower slopes in the coverage results. At $p=10^{-4}$, removing $\sim$30-35\% of Arqade's pipeline stages decreases coverage by only 4.66\%, 2.2\%, and 3.58\%, respectively, for the $[[108,8,10]]$, $d=9$ color, and $[[72,8,10]]$ codes. Coverage loss steepens in the BB and color codes past 35\% stage removal, marking the transition from less important hook-like to more important spacetime-like pipeline stages. Conversely, coverage loss stays shallow in the GB code; this is typical for higher-weight stabilizer codes with deep pipelines, since depth increase is dominated by hook-like stages. Hence, we expect more generally that stage removal is an effective strategy to combat latency constraints for high-weight codes.

In Arqade's ASIC implementations for the $[[108,8,10]]$ and $d=9$ color codes, removing 35\% of pipeline stages translates to $\sim$32\% reduction in area and $\sim$37.63-39.24\% power savings. With additional technology-level optimizations, including voltage scaling, further power reductions are possible, enabling 4 K predecoder support for even more logical qubits.
\section{Discussion \& Related Work} \label{sec:discussion}
\noindent
\textbf{Predecoding:} Many surface code predecoders have been proposed in the past \cite{delfosse2020hierarchical, ravi2023better, smith2023local, alavisamani2024promatch, knapen2026pinball}. To the best of our knowledge, Arqade is the first to extend predecoding beyond the surface code to general qLDPC code architectures.

\noindent
\textbf{Beyond CSS Codes:} The qLDPC codes evaluated in this work are all CSS codes \cite{calderbank1996good}. None of Arqade's design rules are CSS code-specific, so it should generalize to non-CSS codes like the $XZZX$ surface code \cite{bonilla2021xzzx}. Testing the limits of Arqade's generalizability is an interesting direction for future work.

\noindent
\textbf{Decoders:} Other second-level decoders exist for general qLDPC codes, including BP with guided decimation \cite{yao2024belief} and BP with localized statistics decoding \cite{hillmann2025localized}, which trade off latency and accuracy. This work uses BP-OSD and RelayBP-5 due to their proven accuracy over the widest range of qLDPC codes. Arqade passes unmodified syndromes to the second-level, so it is compatible with any decoder.

\noindent
\textbf{SM Circuit Construction:} Recent work has explored optimizing SM circuit construction to reduce logical error rates \cite{viszlai2026prophunt, liu2026alpha}. Evaluating Arqade on circuits generated with these tools should increase its benefits. Future work could explore predecoder-SM circuit co-optimization, such as tailoring error propagation to minimize predecoding primitive conflicts.

\section{Conclusion} \label{sec:conclusion}
This paper introduced Arqade, an automated framework for generating general qLDPC code predecoders. By solving significant fractions of the decoding workload, including scenarios where BP decoders fail to converge, Arqade reduces decoder utilization as well as expensive BP-OSD and RelayBP post-processing. These benefits mitigate system-wide decoding resource contention, reduce average decoding latency, and simplify decoder allocation. The possibility of cryogenic predecoding additionally decreases syndrome transmission bandwidth. Importantly, Arqade realizes these benefits while maintaining state-of-the-art accuracy across qLDPC codes.

\section*{Acknowledgments}
The authors would like to thank Lily Plotner for valuable recommendations regarding detector error models and logical observables as well as Ryan Kersten for insightful discussions on color code constructions.
This material is based upon work supported by the U.S. Department of Energy, Office of Science, Office of Advanced Scientific Computing Research, Accelerated Research in Quantum Computing under Award Number DE-SC0025633. This research used resources of the National Energy Research Scientific Computing Center, a DOE Office of Science User Facility supported by the Office of Science of the U.S. Department of Energy under Contract No. DE-AC02-05CH11231 using NERSC award NERSC DDR-ERCAP0035341.
This research is supported in part by funding from the Quantum Research Institute at the University of Michigan.
The authors also thank Semiwise Ltd., UK for access to the cryo-CMOS PDK used for hardware evaluation in this work.
This research was, in part, funded by the U.S. Government. The views and conclusions contained in this document are those of the authors and should not be interpreted as representing the official policies, either expressed or implied, of the U.S. Government.


\bibliographystyle{IEEEtranS}
\bibliography{refs}

\end{document}